\documentclass[prc,twocolumn,superscriptaddress,showpacs]{revtex4}
\usepackage{graphicx,amsmath,amssymb,bm}

\def\be{\begin{equation}}
\def\ee{\end{equation}}
\def\ba{\begin{eqnarray}}
\def\ea{\end{eqnarray}}
\def\bas{\begin{eqnarray*}}
\def\eas{\end{eqnarray*}}

\newcommand{\la}{\Lambda}
\newcommand{\vlowk}{V_{{\rm low}\,k}}

\newcommand{\fmi}{\, \text{fm}^{-1}}
\newcommand{\mev}{\, \text{MeV}}
\newcommand{\kev}{\, \text{keV}}

\begin{document}

\title{Coupled-cluster theory for three-body Hamiltonians}

\author{G.~Hagen}
\affiliation{Physics Division, Oak Ridge National Laboratory,
P.O. Box 2008, Oak Ridge, TN 37831, USA}
\affiliation{Department of Physics and Astronomy, University of
Tennessee, Knoxville, TN 37996, USA}
\affiliation{Centre of Mathematics for Applications, University of Oslo, 
N-0316 Oslo, Norway} 
\author{T.~Papenbrock}
\affiliation{Physics Division, Oak Ridge National Laboratory,
P.O. Box 2008, Oak Ridge, TN 37831, USA}
\affiliation{Department of Physics and Astronomy, University of
Tennessee, Knoxville, TN 37996, USA}
\author{D.J.~Dean}
\affiliation{Physics Division, Oak Ridge National Laboratory,
P.O. Box 2008, Oak Ridge, TN 37831, USA}
\author{A.~Schwenk}
\affiliation{TRIUMF, 4004 Wesbrook Mall, Vancouver, BC, Canada, V6T 2A3}
\affiliation{Department of Physics, University of Washington,
Seattle, WA 98195, USA}
\author{A.~Nogga}
\affiliation{Institut f\"ur Kernphysik, Forschungszentrum J\"ulich, 
D-52425 J\"ulich, Germany}
\author{M.~W{\l}och}
\affiliation{Department of Chemistry,
Michigan State University, East Lansing, MI 48824, USA}
\author{P.~Piecuch}
\affiliation{Department of Chemistry,
Michigan State University, East Lansing, MI 48824, USA}


\begin{abstract}
We derive coupled-cluster equations for three-body Hamiltonians. The
equations for the one- and two-body cluster amplitudes are presented
in a factorized form that leads to an efficient numerical
implementation. 
We employ low-momentum two- and three-nucleon
interactions and calculate the binding energy of $^4$He. The results
show that the main contribution of the three-nucleon interaction
stems from its density-dependent zero-, one-, and two-body
terms that result from the normal ordering of the Hamiltonian in
coupled-cluster theory. The residual three-body terms that remain
after normal ordering can be neglected.
\end{abstract}


\maketitle

\section{Introduction}

One of the central challenges in nuclear theory is to understand and
predict the structure of nucleonic matter based on microscopic
nucleon-nucleon (NN) and many-nucleon interactions. In recent years,
there has been significant progress in exact calculations of ground
and excited states of light nuclei based on various high-precision 
interactions fitted to NN data~\cite{Kam01,Piep01,Wir02,%
Nog02,Nav00,Nav03,Nog06}. These results clearly show that three-nucleon
forces (3NFs) contribute significantly: Without 3NFs, the binding energies
depend strongly on the NN potential used, which can be traced to scheme 
and model dependences in any theory restricted to NN interactions.
The existence of 3NFs is not surprising, since nucleons are not point 
particles. There are always virtual excitations (high-momentum 
nucleons and $\Delta$-isobars) or internal degrees of freedom (quarks 
and gluons) that have been ``integrated out''. This directly leads to three-
and many-nucleon interactions. The modern understanding is that nuclear 
interactions are effective interactions and depend on the resolution
scale given by a cutoff $\la$ of the effective theory (see for 
example~\cite{Lepage}). Exact calculations are cutoff-independent
up to the effects of omitted higher-order interactions, and 
therefore varying the resolution scale is a powerful tool to analyze
the predictive power of theoretical calculations.

The study of 3NFs in systems beyond the lightest nuclei is an
important goal. This requires a flexible technique to solve the 
many-body problem including NN and 3N interactions. Coupled-cluster 
theory is a promising tool for this endeavor. This method 
originated in nuclear physics~\cite{Coe58,Coe60} 
and is today mostly propelled through its
importance in quantum chemistry~\cite{Ciz66,Ciz69}. For reviews, we
refer the reader to Refs.~\cite{Kuem78,Bart89,Pal99,Craw00,Pie04}. After the
seminal work by the Bochum group~\cite{Kuem78}, Heisenberg and Mihaila
employed coupled-cluster theory for structure calculations of $^{16}$O
based on realistic NN potentials~\cite{Hei99}. For their calculation
of the charge form factor of $^{16}$O, they also included selected
contributions from 3NFs that could be cast into the form of
density-dependent NN interactions~\cite{Mi00}. Another recent
approach employed {\it ab-initio} coupled-cluster theory for 
structure calculations in closed-shell nuclei $^4$He and 
$^{16}$O~\cite{Dean04,Kow04,Wlo05}, in open-shell nuclei as the 
neighbors of $^{16}$O~\cite{Gour06}, and weakly bound and unbound 
helium isotopes~\cite{Hag06}. These calculations were limited to 
NN interactions. It is the purpose of this paper to develop
coupled-cluster theory for three-body Hamiltonians. This extension 
of the coupled-cluster method might also find applications
in condensed-matter theory and quantum chemistry~\cite{Weg94,Glaz94,White02}.

A current frontier in nuclear structure theory is to determine
consistent 3NFs corresponding to the different NN interactions, and
with predictive power when extrapolated to the extremes of isospin and
to moderate densities.  Several theoretical approaches are currently
being used. The Tucson-Melbourne 3NF already employed symmetries of
QCD in its construction~\cite{Coon,Coon93}. Existing phenomenological
3NFs include the Fujita-Miyazawa force based on $2 \pi$ exchange with
an intermediate $\Delta$-isobar~\cite{FM57} and also various
shorter-ranged 3NFs~\cite{Rob86,Ruhrpot,Pud92,Illinois2}.  This
approach has led to a very successful description of light nuclei (for
a review see Ref.~\cite{Piep01}).

Onother approach is to systematically construct NN and higher-order
interactions within the framework of chiral effective field theory
(EFT)~\cite{Bed02,Epel06,N3LO,N3LOEGM}. Chiral interactions are
expanded in powers of a typical momentum of nucleons in nuclei and in
powers of the pion mass, both generically called $Q$, over the EFT
breakdown scale $\Lambda_\chi \sim 1 \, \text{GeV}$ ($\approx 5 \fmi$). 
The EFT power counting naturally explains the hierarchy of NN, 3N, and
higher-body interactions~\cite{Weinberg}, which only enter in
subleading orders and make calculations for complex nuclei based on NN
and 3N interactions meaningful. At this point, the leading 3NF has
been implemented~\cite{ch3NF1,ch3NF2,ch3NF3}.

Nuclear interactions require regularization and
renormalizaton to be meaningful, and with EFT,
usually a momentum cutoff scheme
is used~\cite{Ord96,Epel00,N3LO,N3LOEGM}. The description of the NN
data is not completely independent of the cutoff, but the cutoff
variation decreases with increasing order, because the cutoff
dependences can be absorbed by higher-order contact interactions. In
this way, chiral EFT implements the renormalization group (RG) running
of the interactions up to higher-order terms. This is reflected in the
leading chiral 3NF, which includes a 3N contact term to be fitted to
3N data. In principle, nuclear structure calculations based directly
on chiral interactions are
feasible~\cite{Nog06,NavCau,NavEFT1,NavEFT2}. However, the typical cutoffs
($\Lambda \approx 2.5-3 \fmi$) are somewhat too large to use
chiral interactions without resummations or prediagonalizations in
many-body calculations.

Renormalization group methods can be used to evolve nuclear 
interactions to lower momenta, which leads to improved convergence
in few- and many-body calculations~\cite{Vlowk1,Vlowk2,Vlowk3N,nucmatt}.
The resulting low-momentum interactions, known generically as 
``$V_{{\rm low}\,k}$'', have variable momentum cutoffs and are
approximately independent of the starting NN interaction for
cutoffs $\la \lesssim 2 \fmi$~\cite{Vlowk1,Vlowk2}. The RG
evolution preserves the long-range parts, and starting from
chiral EFT interactions, generates all higher-order contact 
operators needed to reproduce low-energy NN observables.
Moreover, with increasing orders in EFT, the resulting 
$\vlowk$ interactions are very similar to low-momentum interactions 
obtained from conventional potentials~\cite{Vlowk2,Schw05}.
Since chiral EFT represents the most general low-momentum expansion
of nuclear forces, the above observations motivate combining
low-momentum interactions with 3NFs from chiral
EFT~\cite{Vlowk3N}. We will follow this approach and
employ $\vlowk$ with the corresponding low-momentum 3NF
adjusted to the binding energies of $^3$H and $^4$He~\cite{Vlowk3N}.
In this way, we can expect to define approximately consistent
NN and 3N interactions. In this first study, we focus on the
development of the coupled-cluster method to include 3NFs and will 
only present results for one cutoff. A study of the cutoff variation
and associated uncertainties due to higher-order many-body interactions
will be left to future work.

This paper is organized as follows. In Section~\ref{derive} we present
coupled-cluster theory and its extension to three-body Hamiltonians.
In Section~\ref{results}, we solve the resulting coupled-cluster
amplitude equations at the singles and doubles (CCSD) level for $^4$He
based on low-momentum NN and 3N interactions.  Our findings are especially
promising as they show that the binding energy of $^4$He can be
calculated based on those parts of the 3NF that can be viewed as
density-dependent zero-, one-, and two-body forces. We summarize our results 
in Section~\ref{summary}.

\section{Coupled-cluster equations for three-nucleon forces}
\label{derive}

This is the main technical section of this paper. In the first
subsection, we briefly recapitulate coupled-cluster theory and the
reformulation of the three-body Hamiltonian in normal-ordered
form. The second subsection deals with the diagrammatic derivation and
the factorization of the coupled-cluster equations due to the residual 
three-body force that remains after normal ordering of the Hamiltonian.

\subsection{Coupled-cluster theory}

We consider a pure three-body Hamiltonian. Coupled-cluster theory for
one- and two-body Hamiltonians
is a mature field, and we refer the reader to
the reviews~\cite{Kuem78,Bart89,Pal99,Craw00,Pie04}. 
The three-body Hamiltonian is written as
\be
\label{ham3}
\hat{H}_3= {1\over 36} \sum_{pqrstu} \langle pqr||stu\rangle 
\hat{a}^\dagger_p \hat{a}^\dagger_q \hat{a}^\dagger_r \hat{a}_u \hat{a}_t 
\hat{a}_s \,.
\ee
Here, $\langle pqr||stu\rangle$ denotes the antisymmetrized three-body 
matrix elements, while $\hat{a}^\dagger_p$ and $\hat{a}_p$ create and 
annihilate a fermion
in the single-particle orbital $p$, respectively. 

In coupled-cluster theory, the Fermi vacuum is a single-particle
product state $|\phi\rangle=\prod_{i=1}^A \hat{a}_i^\dagger|0\rangle$,
where the $A$ lowest-energy orbitals are occupied. In a first step, we
cast the three-body interaction into a normal-ordered form with
respect to this vacuum.  In what follows it is assumed that the
indices $i, j, k, l, m$ label the occupied orbitals of $|\phi\rangle$
while $a, b, c, d, e$ refer to the unoccupied orbitals of
$|\phi\rangle$. The former indices run over the number $n_o\equiv A$
of occupied orbitals, while the latter run over the remaining number
$n_u$ of unoccupied orbitals. Typically, one has $n_u\gg n_o$. Indices
referring to all orbitals are denoted as $p, q, r, s, t, u$; see for
example Eq.~(\ref{ham3}).  The normal-ordered Hamiltonian is thus
\ba
\label{normal}
\hat{H}_3 &=& {1\over 6} \sum_{ijk} \langle ijk||ijk\rangle 
+ {1\over 2} \sum_{ijpq} \langle ijp||ijq\rangle 
\{\hat{a}^\dagger_p \hat{a}_q \} \\\nonumber
&& + {1\over 4} \sum_{ipqrs} \langle ipq||irs\rangle 
\{\hat{a}^\dagger_p \hat{a}^\dagger_q \hat{a}_s \hat{a}_r \} + \hat{h}_3 \,,
\ea
where $\hat{h}_3$ denotes the residual three-body Hamiltonian 
\be
\label{h3}
\hat{h}_3\equiv {1\over 36} \sum_{pqrstu} \langle pqr||stu\rangle 
\{\hat{a}^\dagger_p \hat{a}^\dagger_q \hat{a}^\dagger_r \hat{a}_u \hat{a}_t 
\hat{a}_s \} \,.
\ee
Here, we used the $\{\ldots\}$ to denote normal ordering.  
Writing the three-body Hamiltonian in normal-ordered form, it is clear
that the Hamiltonian separates into a zero-, one-, two-, and a three-body
term. The first three sums in Eq.~(\ref{normal}) are the vacuum
expectation value, and the ``density-dependent'' one- and two-body
terms, respectively. Their treatment is standard in coupled-cluster
theory, as they simply modify the normal-ordered two-body Hamiltonian
from the $NN$ interaction.  In this section, we focus on the residual
three-body operator of Eq.~(\ref{h3}).

In coupled-cluster theory, the correlated state $|\psi\rangle$ is given
by a correlation operator $\exp{(\hat{T})}$ that acts onto a
single-particle product state $|\phi\rangle=\prod_{i=1}^A
\hat{a}_i^\dagger|0\rangle$ of the $A$-body system by
\be
\label{ansatz}
|\psi\rangle=e^{\hat{T}}|\phi\rangle \,.
\ee
The cluster operator,
\be
\label{t}
\hat{T}=\hat{T}_1 + \hat{T}_2 + \ldots + \hat{T}_A \,,
\ee
consists of a one-body cluster operator
\be
\label{t1}
\hat{T_1}= \sum_{i a} t_i^a \hat{a}_a^\dagger \hat{a}_i \,,
\ee
a two-body cluster operator
\be
\label{t2}
\hat{T_2}= {1\over 4}\sum_{i j a b} t_{ij}^{ab} \hat{a}^\dagger_a 
\hat{a}^\dagger_b \hat{a}_j \hat{a}_i \,,
\ee
and so forth. Note that the two-body cluster amplitudes
$t_{ij}^{ab}=-t_{ij}^{ba}=-t_{ji}^{ab}=t_{ji}^{ba}$ are fully
antisymmetric. Clearly, $\hat{T}_1$ and
$\hat{T}_2$ induce $1p-1h$ and $2p-2h$ excitations, respectively.  In
what follows, we will limit the expansion, Eq.~(\ref{t}), of the cluster
operator to the one-body cluster, Eq.~(\ref{t1}), and the two-body cluster,
Eq.~(\ref{t2}), respectively. This approximation is referred to as CCSD
(``coupled-cluster theory with single and double excitations''). CCSD
is a powerful approximation and a compromise between accuracy
on the one hand and computational effort on the other hand. One inserts
the ansatz Eq.~(\ref{ansatz}) into the Schr\"odinger equation, 
multiplies with $\exp{(-\hat{T})}$ from the left, and obtains 
the following set of equations
\ba
\label{ccsd}
E &=& \langle \phi | \overline{H} | \phi\rangle \,, \\
\label{ccsd1}
0 &=& \langle \phi_i^a | \overline{H} | \phi\rangle \,, \\
\label{ccsd2}
0 &=& \langle \phi_{ij}^{ab} | \overline{H} | \phi\rangle \,.
\ea
Here $|\phi_{i_1\ldots i_n}^{a_1\ldots a_n}\rangle = 
\hat{a}_{a_n}^\dagger\ldots \hat{a}_{a_1}^\dagger 
\hat{a}_{i_1}\ldots \hat{a}_{i_n}|\phi\rangle$ is a $np-nh$ excitation of the 
product state $|\phi\rangle$, and 
\be
\label{hsim}
\overline{H} = \exp{(-\hat{T})} \hat{H} \exp{(\hat{T})}
\ee 
is the similarity-transformed Hamiltonian. This Hamiltonian is a sum
of one-, two-, and three-body Hamiltonians, i.e.
$\hat{H}=\hat{H}_1+\hat{H}_2+\hat{H}_3$. The treatment of this
Hamiltonian within coupled-cluster theory is well known, 
except for the residual three-body term,
Eq~(\ref{h3}), of the normal-ordered three-body Hamiltonian $\hat{H}_3$. 

The CCSD Eqs.~(\ref{ccsd1}) and~(\ref{ccsd2}) determine 
the one-particle and two-particle cluster amplitudes $t_i^a$ and
$t_{ij}^{ab}$, respectively. These amplitudes can then be inserted
into the first of the CCSD equations, Eq.~(\ref{ccsd}), to determine the
ground-state energy.  Note that the similarity-transformed Hamiltonian
of Eq.~(\ref{hsim}) is not Hermitian, and CCSD is not a variational
approach. However, the similarity-transformed Hamiltonian can be 
evaluated {\it exactly} for any truncation of the cluster operator. 
In what follows, we will compute the corrections to the energy,
Eq.~(\ref{ccsd}), and to the CCSD Eqs.~(\ref{ccsd1}) 
and (\ref{ccsd2}), that arise due to the residual three-body
Hamiltonian.

\subsection{Derivation of coupled-cluster equations}
\label{maintheo}

In this subsection, we derive the contribution of the residual
three-body term $\hat{h}_3$ from Eq.~(\ref{h3}) to the energy,
Eq.~(\ref{ccsd}), and the cluster amplitudes, Eqs.~(\ref{ccsd1}) 
and~(\ref{ccsd2}). The final results are given in Eq.~(\ref{erg}) for the
energy and in Eqs.~(\ref{t1fac}) and~(\ref{t2fac}) for the
cluster amplitudes, respectively. The contributions to the energy and
the coupled-cluster amplitudes can be derived, for instance, by following 
the approach of Ref.~\cite{Craw00}. The resulting expressions contain 
2, 15 and 51 terms for the energy and the cluster amplitudes. Many 
of the individual terms consist of sub-terms of similar structure. 
The theoretical derivation presented in this subsection exploits this 
structure and leads to an efficient numerical implementation.

It is most convenient to evaluate the matrix elements in
Eq.~(\ref{ccsd}) of the similarity-transformed Hamiltonian, 
Eq.~(\ref{hsim}), in a diagrammatic
form (see, for example,~\cite{Kuch86,Craw00}). Diagrams are a useful
book keeping device to keep track of the (considerable) number of
possible Wick contractions.  We refer the reader to the
literature for a more detailed description.

\begin{figure}[t]
\includegraphics[width=0.26\textwidth,angle=0]{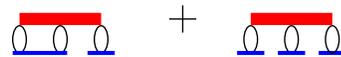}
\caption{(Color online) Energy contributions of the residual three-body 
Hamiltonian, Eq.~(\ref{h3}), in the CCSD approximation.}
\label{fig1}
\end{figure}

In a first step, we determine the correction to the CCSD energy,
Eq.~(\ref{ccsd}), that is due to the residual three-body
Hamiltonian~(\ref{h3}).  The matrix element $\langle \phi |
\overline{h_3} |\phi\rangle$ is a sum of all topologically different
diagrams, where the Hamiltonian, Eq.~(\ref{h3}), is fully contracted by
the cluster operators.  The two diagrams that enter this expression
are presented in Fig.~\ref{fig1}. In these diagrams, the thick
horizontal bar represents the residual three-body Hamiltonian,
Eq.~(\ref{h3}); the thin horizontal bars denote the one-body and
two-body cluster operators, respectively. Particle and hole lines are
also shown. If uncontracted, the former have an arrow that points
upward while the latter have an arrow that points downward. The
corresponding algebraic expression for the energy correction is
\ba
\label{erg}
e_3 &=& \langle\phi|\bigl(\hat{H}\hat{T}_1\hat{T}_2\bigr)_c|\phi\rangle 
+ \langle\phi|\bigl({1\over 6} 
\hat{H}\hat{T}_1^3\bigr)_c|\phi\rangle \,, \\\nonumber
&=& {1\over 4} \sum_{klmcde}\langle klm||cde\rangle t_k^c t_{lm}^{de}
\\\nonumber
&+& {1\over 6} \sum_{klmcde}\langle klm||cde\rangle t_k^c t_l^d t_m^e \,. 
\ea
This is the energy correction due to the residual three-body Hamiltonian. 
Note that the
computational effort of the energy, Eq.~(\ref{erg}), scales as $O(n_u^3
n_o^3)$. Note also that the summation over $\sum_{kc} t_k^c$ is common
to both diagrams and might therefore be factored out. This is the
basic idea behind the factorization~\cite{Kuch86}, and will be
presented in detail below.

\begin{figure}[t]
\includegraphics[width=0.36\textwidth,angle=0]{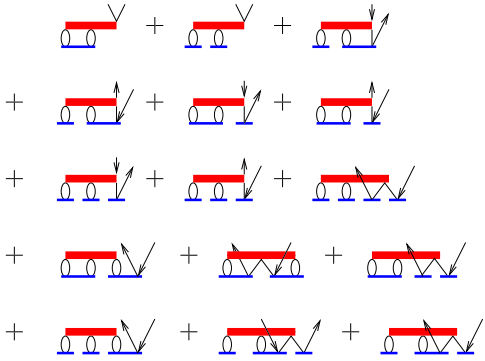}
\caption{(Color online) 
Contributions of the three-body Hamiltonian, Eq.~(\ref{h3}),
to the $\hat{T}_1$ cluster equation in the CCSD approximation.}
\label{fig2}
\end{figure}

Let us consider the matrix element $\langle \phi_i^a |
\overline{H} | \phi\rangle$ appearing in the CCSD Eq.~(\ref{ccsd1}). 
It is given by the sum over all topologically different
diagrams where one particle line and one hole line are not contracted.
Figure~\ref{fig2} shows the 15 diagrams that enter this matrix
element. 

The translation of these diagrams into algebraic expressions is
straightforward, but there are two reasons to not present them here.
First, the naive numerical implementation of the resulting expressions
would be inefficient. To see this, consider, for instance, the third 
diagram in Fig.~\ref{fig2}.  The corresponding algebraic expression is
\be
-{1\over 2}\sum_{ckdlm}\langle klm|| cdi\rangle t_k^c t_{lm}^{da} \,,
\ee
and its naive evaluation costs $O(n_u^3n_o^4)$ operations. However,
performing the summations involving the $\hat{T}_1$ cluster operator
first yields the intermediate
\be
\label{Iex}
I_{di}^{lm} \equiv {1\over 2}\sum_{ck}\langle klm|| cdi\rangle t_{k}^{c} \,,
\ee
which only costs $O(n_u^2n_o^4)$ operations and requires $O(n_u
n_o^3)$ in memory.  The subsequent contraction of this intermediate
with the remaining $\hat{T}_2$ cluster operator only costs
$O(n_u^2n_o^3)$ operations.  Clearly, the memory cost of the
intermediate is overcompensated by the reduction of computational
cycles. Note also that the intermediate Eq.~(\ref{Iex}) enters in
the evaluation of the seventh diagram depicted in Fig.~\ref{fig2}. The
second reason is that the complexity of the involved diagrams
increases rapidly. The number of diagrams increases from 2 to 15 to
51, when going from Eq.~(\ref{ccsd}) to Eq.~(\ref{ccsd1}) to
Eq.~(\ref{ccsd2}), respectively. The construction of each individual
diagram and its inspection regarding the construction and use of
intermediates then becomes cumbersome, and a more systematic approach
is called for. Similar comments apply when improving the coupled-cluster 
wave function through the inclusion of three-body or four-body
cluster amplitudes.  One therefore considers a factorization of the
coupled-cluster equations~\cite{Kuch86}. This approach yields a very
compact form of the coupled-cluster equations and is particularly
useful for the numerical implementation~\cite{Kuch91,Kuch92,Pie02}. So
far, factorized coupled-cluster equations have been derived in a
two-step procedure.  The first step consists of constructing all
topologically different coupled-cluster diagrams. In a second step,
these diagrams are analyzed and repeatedly decomposed into simpler
intermediates that undergo single contractions. Here, we proceed
differently and present a {\it direct} diagrammatic derivation of the
factorized coupled-cluster equations. Our derivation avoids the
explicit construction of all individual coupled-cluster diagrams.

\begin{figure}[t]
\includegraphics[width=0.46\textwidth,angle=0]{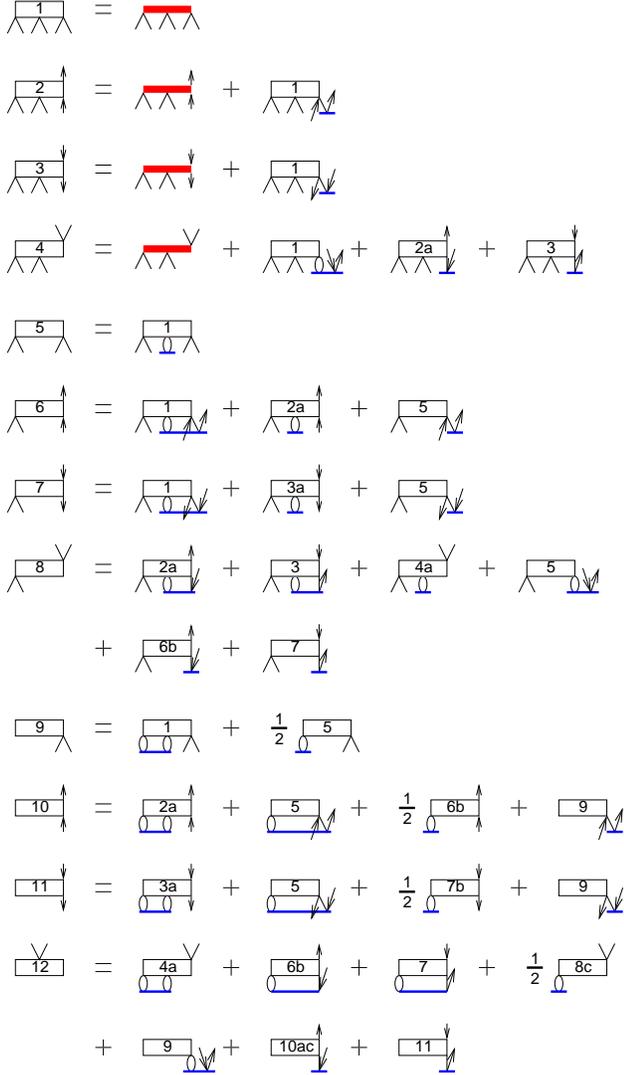}
\caption{(Color online) Matrix elements of the residual three-body 
part of the similarity-transformed Hamiltonian, Eq.~(\ref{hsim}), 
that enter the amplitude Eq.~(\ref{ccsd1}).}
\label{fig3}
\end{figure}

\begin{table*}
\caption{Memory multiplication factor $m$ and computational cost 
factor $c$ for forming intermediates through
contraction with the $k$-body cluster operator $T_k(p^n h^m)$ 
via $n$ particle lines and $m$ hole lines.}
\begin{center}
\begin{tabular}{|c|c|c|c|c|c|c|c|c|c|c|c|}
\hline
 &$ T_2(p^2h^2) $&$ T_1(ph) $&$ T_2(p^2h) $&$ T_1(p) $&$ T_2(p^2) $&
$ T_2(ph^2) $&$ T_2(ph)$&$ T_2(p) $&$ T_2(h^2) $&$ T_1(h) $&$ T_2(h)$ \cr\hline
$m$&$(n_u n_o)^{-2} $&$ (n_u n_o)^{-1} $&$ n_u^{-2} $&$ n_o/n_u $&
$ (n_o/n_u)^2 $&$ n_o^{-2} $&$ 1 $&$ n_o^2 $&$ (n_u/n_o)^2 $&$ n_u/n_o $&
$ n_u^2$ \cr\hline
$c$&$ 1 $&$ 1 $&$ n_o $&$ n_o $&$ n_o^2 $&$ n_u $&$ n_un_o $&$ n_u n_o^2 $
&$ n_u^2 $&$ n_u $&$ n_u^2n_o$ \cr\hline
\end{tabular}
\end{center}
\label{tab1}
\end{table*}

We have to decide in which order multiple contractions of the
Hamiltonian with the cluster operators should be performed. Let  
$T_k(p^n h^m)$ denote the contraction of the Hamiltonian with the 
$k$-body cluster
operator $\hat{T}_k$ via $n$ particle lines and $m$ hole lines.
The contraction of $T_k(p^n h^m)$ with an object of $i$ particle lines
and $j$ hole lines costs $c \, n_u^in_o^j$ computational operations and 
results in an object of size $m \, n_u^in_o^j$. Here, $c$ and $m$ denote
the computational cost and memory multiplier, respectively, and one finds
\bas
c &=& n_u^{k-n}n_o^{k-m} \,, \\[2mm]\nonumber
m &=& n_u^{k-2n}n_o^{k-2m} \,. \nonumber
\eas
Based on this analysis, we find that the cost of two subsequent
contractions labeled $T_A$ and $T_B$, respectively, is proportional to
$c(T_A) + m(T_A) c(T_B)$ when contraction $T_A$ is first, and
proportional to $c(T_B) + m(T_B) c(T_A)$ when contraction $T_B$ is
first. Table~\ref{tab1} shows (from left to right) the optimal order
in which subsequent contractions should be performed, under the
condition that $1\ll n_o\ll n_u$ and $n_o^2<n_u$. We also listed the
relative computational cost $c$, and the memory multiplier $m$. For
the first 7 entries, the order is easily understood.  These
contractions do not yield an increase of the size of the contracted
object (since $m \leqslant 1$), and the order is therefore determined by the
computational cost. The remaining four contractions increase the size
of the contracted object (since $m>1$), and it is usually most
efficient to perform the computationally more expensive contraction
before performing the second contraction on an object with increased
size.

Let us now turn to the diagrammatic factorization of the
coupled-cluster equations. Considering Eq.~(\ref{ccsd1}) and
Fig.~\ref{fig2}, we have to construct all coupled-cluster diagrams
with one incoming hole line and one outgoing particle line. These
diagrams should be constructed from simpler diagrams, adding one
contraction at each step. For the residual three-body Hamiltonian, the
root is clearly given by the diagram 1 in Fig.~\ref{fig3}. This
diagram is the only one that has three incoming particle lines and
three outgoing hole lines. Diagrams 2 and 3 have one outgoing particle
line and one incoming hole line, respectively, and also have a total
of six outgoing and incoming lines. They are sums of two diagrams. The
first is the residual three-body Hamiltonian with this appropriate
particle and hole lines, while the second diagram is a contraction of
diagram 1 with a $\hat{T}_1$ cluster operator.

In what follows, we adopt the following convention. For a diagram that
is the sum of diagrams, we label the first, second, third, etc. term
of the sum by a, b, c, respectively. For example, diagram 2 is the sum
of diagram 2a and diagram 2b. These labels are not
printed in Fig.~\ref{fig3}. Diagram 4 is the sum of 4 diagrams, namely
the corresponding residual three-body Hamiltonian and three contractions of
previously generated diagrams. Note that out of the two diagrams of
diagram 2, only diagram 2a enters. This is due to the order specified
in Table~\ref{tab1}. Diagram 2b is a contraction of diagram 1 with
$T_1(p)$, and this contraction cannot be followed by the contraction
$T_1(p)$. It is now clear how to proceed. Diagrams 5 to 8 have a total
of four incoming and outgoing lines, while diagrams 9 to 12 have a
total of two incoming and outgoing lines, respectively. Note the
peculiar factor $1/2$ in front of diagram 9b. This diagram is a
$T_1(ph)$ contraction of diagram 5b which itself is also a $T_1(ph)$
contraction. The factor $1/2$ will be needed for the translation into
algebraic expressions. Similar comments apply to diagrams 10c, 11c,
and 12d. Note that diagram 12 consists of all diagrams with one
incoming hole line and one outgoing particle line. The recursive 
expansion of the corresponding right-hand side yields indeed all 
diagrams depicted in Fig.~\ref{fig2}, and therefore factors the CCSD
Eq.~(\ref{ccsd1}). 

\begin{figure*}[t]
\includegraphics[width=0.48\textwidth,angle=0]{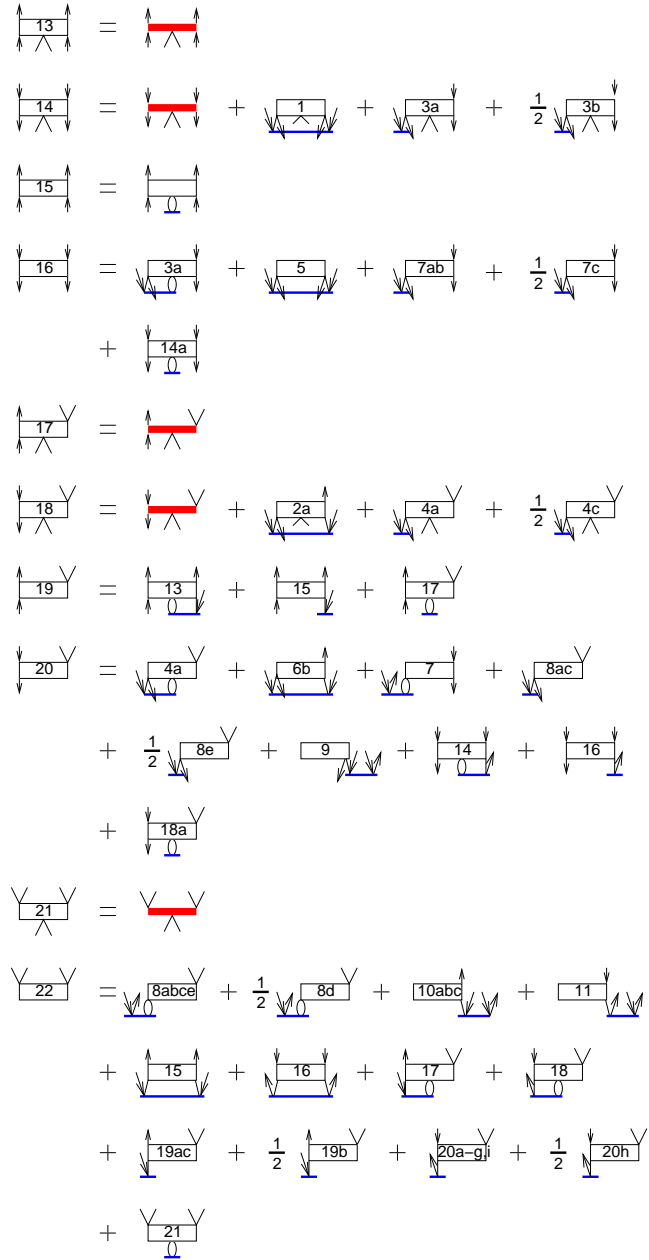}
\caption{(Color online) Additional diagrams needed for the
contributions of the residual three-body Hamiltonian, Eq.~(\ref{h3}), 
to the CCSD Eq.~(\ref{ccsd2}).}
\label{fig4}
\end{figure*}

Let us translate Fig.~\ref{fig2} into algebraic expressions. We use
the convention that the intermediate $I_{p_1,p_2,\ldots}^{q_1,
q_2,\ldots}$ has incoming lines $p_1, p_2, \ldots$ and outgoing lines
$q_1, q_2, \ldots$, respectively. In what follows, the intermediate
$I(\nu)$ corresponds to diagram $\nu $ of Fig.~\ref{fig3}. We
restrict ourselves to those intermediates that are needed for the
construction of diagram 12, and denote the diagrams 1, 2a, 3a, and 4a
directly in terms of the corresponding three-body matrix elements.
The result is
\bas
I(5)_{ce}^{km} &=& {1\over 4}\sum_{dl}\langle klm||cde\rangle t_l^d \,,
\\\nonumber
I({\rm 6b})_{ce}^{ka} &=& {1\over 2}\sum_{dl}\langle kla||cde\rangle t_l^d \,,
\\\nonumber
I({\rm 7b})_{ci}^{km} &=& {1\over 2}\sum_{dl}\langle klm||cdi\rangle t_l^d \,,
\\\nonumber
I(7)_{ci}^{km} &=& {1\over 4}\sum_{lde}\langle klm||cde\rangle t_{li}^{de} 
+ I({\rm 7b})_{ci}^{km} \\\nonumber
&+& 2\sum_eI(5)_{ce}^{km}t_i^e \,,\\\nonumber
I({\rm 8c})_{ci}^{ka} &=& \sum_{dl}\langle kla|| cdi\rangle t_l^d \,,
\\\nonumber
I(9)_c^k &=& {1\over 4}\sum_{delm} \langle klm||cde\rangle t_{lm}^{de} 
+ 2\sum_{em}I(5)_{ce}^{km} t_m^e \,, \nonumber
\eas
and
\bas
I({\rm 10ac})_e^a &=& {1\over 4}\sum_{ckdl}\langle kla||cde\rangle t_{kl}^{cd} 
+ \sum_{kc}I({\rm 6b})_{ce}^{ka} t_k^c \,, \\\nonumber
I(11)_i^m &=& {1\over 4}\sum_{ckdl} \langle klm||cdi\rangle t_{kl}^{cd} 
+ 2 \sum_{kce}I(5)_{ce}^{km} t_{ki}^{ce} \\\nonumber
&+& \sum_{kc}I({\rm 7b})_{ci}^{km} t_k^c +\sum_c I(9)_c^m t_i^c \,.\nonumber
\eas
In terms of these intermediates, the correction to the right hand side 
of the CCSD Eq.~(\ref{ccsd1}) reads
\ba
\label{t1fac}
&&{1\over 4}\sum_{ckdl}\langle kla||cdi\rangle t_{kl}^{cd} 
+ \sum_{cke}I({\rm 6b})_{ce}^{ka} t_{ki}^{ce} \\\nonumber
&-& \sum_{ckm}I(7)_{ci}^{km} t_{km}^{ca} 
+ {1\over 2}\sum_{kc}I({\rm 8c})_{ci}^{ka} t_k^c 
+ \sum_{ck} I(9)_c^k t_{ki}^{ca} \\\nonumber
&+& \sum_e I({\rm 10ac})_e^a t_i^e - \sum_m I(11)_i^m t_m^a \ .
\ea
These terms have to be added to the well-known CCSD equation for 
the $T_1$-cluster amplitudes based on two-body Hamiltonians.  

For the factorization of the CCSD Eq.~(\ref{ccsd2}), one has to
construct all diagrams with two incoming hole lines and two outgoing
particle lines.  Their number is 51, and we directly 
derive them in a factorized form.  Figure~\ref{fig4} shows the
factorized diagrams that are needed for the recursive construction of
the CCSD Eq.~(\ref{ccsd2}). 
The corresponding algebraic expressions for the intermediates are
\bas
I(3b)^{klm}_{cdi}&=&{1\over 12} \sum_e\langle klm||cde\rangle t_i^e \,,
\\\nonumber
I(3)^{klm}_{cdi} &=&{1\over 12} \langle klm||cdi\rangle + I(3b)^{klm}_{cdi} \,,
\\\nonumber
I(4c)_{cdi}^{kla}&=& {1\over 4} \sum_e\langle kla||cde\rangle t_i^e\,,
\\\nonumber
I(7c)_{ci}^{km}&=& 2\sum_e I(5)_{ce}^{km} t_i^e \,,\\\nonumber
I(7ab)_{ci}^{km}&=& I(7)_{ci}^{km}-I(7c)_{ci}^{km} \,,\\\nonumber
I(8ac)_{ci}^{ka} &=& {1\over 2}\sum_{lde} \langle kla||cde\rangle t_{li}^{de}
+I(8c)_{ci}^{ka} \,,\\\nonumber
I(8b)_{ci}^{ka} &=& -6\sum_{lmd} I(3)_{cdi}^{klm} t_{lm}^{da} \,,\\\nonumber
I(8d)_{ci}^{ka} &=& 4\sum_{me}I(5)_{ce}^{km} t_{mi}^{ea} \,,\\\nonumber
I(8e)_{ci}^{ka} &=& 2\sum_e I(6b)_{ce}^{ka} t_i^e \,,\\\nonumber
I(8abce)_{ci}^{ka} &=& I(8ac)_{ci}^{ka} +I(8b)_{ci}^{ka} +I(8e)_{ci}^{ka} \,, 
\\\nonumber
I(10abc)_e^a &=& I(10ac)_e^a - 2\sum_{kmc} I(5)_{ce}^{km} t_{km}^{ca} \,,
\\\nonumber
I(14)_{idj}^{klm} &=& {1\over 12} \langle klm||idj\rangle 
+ {1\over 24} \sum_{ce}\langle klm||cde\rangle t_{ij}^{ce}\\\nonumber
&+& \sum_c \left({1\over 6} \langle klm ||cdj\rangle 
+ I(3b)_{cdj}^{klm}\right)t_i^c \,,\\\nonumber
\eas
and
\bas
I(15)_{ce}^{ab} &=& {1\over 2} \sum_{ld} \langle alb||cde\rangle t_l^d \,,
\\\nonumber
I(16)_{ij}^{km} &=& {1\over 4} \sum_{cdl} \langle klm||cdj\rangle t_{li}^{dc}
+{1\over 2} \sum_{ce} I(5)_{ce}^{km} t_{ij}^{ce} \\\nonumber
&+& \sum_c\left( I(7ab)_{cj}^{km} + {1\over 2} I(7c)_{cj}^{km}\right)t_i^c
\\\nonumber
&+& {1\over 4} \sum_{ld} \langle klm||idj\rangle t_l^d \,,\\\nonumber
I(18)_{idj}^{klb} &=& {1\over 4} \langle klb||idj\rangle
+ {1\over 8} \sum_{ce} \langle klb||cde\rangle t_{ij}^{ce}\\\nonumber
&+& \sum_c\left({1\over 2}\langle klb||cdj\rangle+I(4c)_{cdj}^{klb}\right)t_i^c
\,,\\\nonumber
I(19ac)_{cj}^{ab} &=& {1\over 2} \sum_{lde} \langle alb||cde\rangle t_{lj}^{de}
+ \sum_{ld} \langle alb||cdj\rangle t_l^d \,,\\\nonumber
I(19b)_{cj}^{ab} &=& 2\sum_e I(15)_{ce}^{ab} t_j^e \,,\\\nonumber
I(20a-g,i)_{ij}^{kb} &=& {1\over 2} \sum_{cld} \langle klb||cdj\rangle 
t_{il}^{cd} +{1\over 2} \sum_{ec} I(6b)_{ce}^{kb} t_{ij}^{ce}\\\nonumber
&+& 2\sum_{mc} I(7)_{ci}^{mk} t_{mj}^{cb} 
+ {1\over 2}\sum_{ld} \langle klb||idj\rangle t_l^d \\\nonumber
&+&\sum_c\left(I(8ac)_{cj}^{kb}+I(8e)_{cj}^{kb}\right) t_i^c\\\nonumber
&+& {1\over 2} \sum_c I(9)_c^k t_{ij}^{cb} 
-3\sum_{ldm} I(14)_{idj}^{klm} t_{lm}^{db} \,,\\\nonumber
I(20h)_{ij}^{kb} &=& -2\sum_m I(16)_{ij}^{km} t_m^b \,.\nonumber
\eas

The residual three-body Hamiltonian, Eq.~(\ref{h3}), thus leads to the 
following correction, to be added to the right hand side of 
the CCSD Eq.~(\ref{ccsd2}),
\begin{widetext}
\ba
\label{t2fac}
&&\sum_{ld}\langle alb||idj\rangle t_l^d
+ {1\over 2} P(ij)\sum_{lcd} \langle alb||cdj\rangle t_{li}^{dc}
+ P(ab)P(ij) \sum_{kc} \left(I(8abce)_{ci}^{ka} 
+ {1\over 2} I(8d)_{ci}^{ka}\right) t_{kj}^{cb} \\\nonumber
&+& P(ab)\sum_e I(10abc)_e^a t_{ij}^{eb}
- P(ij) \sum_m I(11)_i^m t_{mj}^{ab}
+ \sum_{ce}I(15)_{ce}^{ab} t_{ij}^{ce}
+ P(ij)\sum_{km} I(16)_{ij}^{km} t_{km}^{ab} \\\nonumber
&-& P(ab)P(ij)\sum_{kld}I(18)_{idj}^{klb} t_{lk}^{da}
+ P(ij)\sum_c\left(I(19ac)_{cj}^{ab} 
+ {1\over 2} I(19b)_{cj}^{ab}\right)t_i^c \\\nonumber
&-& P(ab)P(ij) \sum_k\left(I(20a-g,i)_{ij}^{kb}
+{1\over 2} I(20h)_{ij}^{kb}\right) t_k^a \,.
\ea
\end{widetext}
Here the permutation $P(ab)$ implies $ P(ab) I_{ab}=I_{ab}-I_{ba}$.
Again, these have to be added to the right hand side of the CCSD 
Eq.~(\ref{ccsd2}). Equation~(\ref{erg}) for the energy correction, and 
the expressions of Eqs.~(\ref{t1fac}) and~(\ref{t2fac}) for the 
cluster amplitudes are the main technical results of
this paper. Upon expansion of diagram 22 in Fig.~\ref{fig4}, one gets
indeed all 51 diagrams that enter the coupled-cluster
Eq.~(\ref{ccsd2}). The numerical implementation of the terms
in Eqs.~(\ref{t1fac}) and~(\ref{t2fac}) is straightforward. 
It is interesting to
analyze the resulting computational costs. The most expensive
intermediates $I(18)$ and $I(19ac)$ cost $n_u^4 n_o^4$ and $n_u^5
n_o^2$ computational cycles, respectively. The most memory-expensive
object is the interaction $\langle alb||cde\rangle$ which enters the
construction of intermediates $I(15)$ and $I(19ac)$, and requires the
storage of $n_u^5n_o$ real numbers. In our largest calculations for $^4$He, 
we have $n_o=4$ and $n_u=220$. 

\section{Application to $^4{\rm He}$}
\label{results}

We present the first {\it ab-initio} coupled-cluster calculations 
including 3NFs. Our results are based on low-momentum NN~\cite{Vlowk2} 
and 3N~\cite{Vlowk3N} interactions,
\be
H = T + \vlowk(\la) + V_{\rm 3N}(\la) \,.
\ee
In this exploratory study we use a sharp cutoff $\la = 1.9 \fmi$,
and $\vlowk$ is derived from the Argonne $v_{18}$ potential~\cite{AV18}.
The corresponding 3N interaction is based on the leading chiral 3NF
and has been fitted to the $^3$H and $^4$He binding energies in 
Ref.~\cite{Vlowk3N}. This 3NF consists of a long-range 
$2\pi$-exchange part, determined by the low-energy coefficients $c_1$,
$c_3$, and $c_4$, an intermediate-range $1\pi$-exchange (``$D$-term'') 
and a short-range contact interaction (``$E$-term'')~\cite{ch3NF1,ch3NF3}.
The 3NF operators are multiplied by regulating functions of the incoming 
and outgoing Jacobi momenta, $f_{\text{R}}(p,q) = \exp\bigl[ - \bigl(
(p^2+ \frac{3 q^2}{4})/\la^2\bigr)^4 \bigr]$, with the same cutoff
value $\la$ as in $\vlowk$. For additional details on the 3NF, we 
refer the reader to Refs.~\cite{Vlowk3N,nucmatt}.
These low-momentum interactions can be directly employed within 
coupled-cluster theory.

\begin{figure}[t]
\includegraphics[width=0.46\textwidth,angle=0]{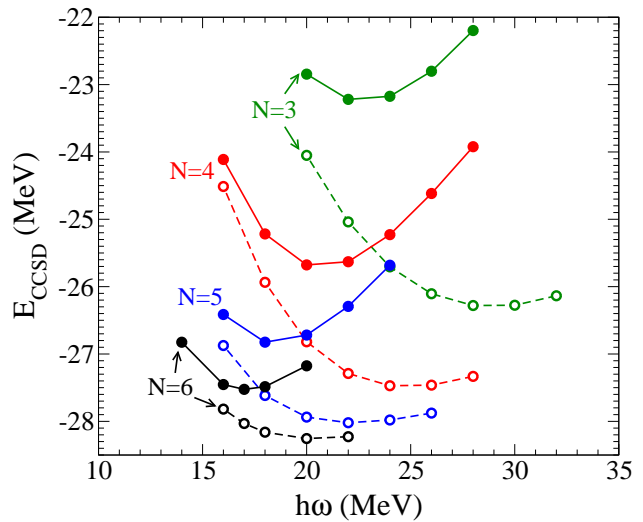}
\caption{(Color online) CCSD results for the binding energy of
$^4$He as a function of the oscillator spacing and for model 
spaces consisting of $N=3$ to $N=6$ oscillator shells.
The CCSD calculations are based on low-momentum NN and 3N 
interactions, where the full and dashed lines respectively
denote the energy obtained with and without 3NFs.}
\label{fig5}
\end{figure}

In this application of coupled-cluster theory we restrict ourselves
to a proof-of-principle calculation. For simplicity, we have therefore
considered only the 3NF channel 
with total isospin $T=1/2$, total angular momentum
$J=1/2$, and positive parity. This partial wave is the dominant
contribution to the binding energies of light nuclei: For $^4$He,
the corresponding Faddeev-Yakubovsky result is $E = - 28.20(5) \mev$,
which differs only by $100 \kev$ from the exact energy including all
partial waves, $E = -28.30(5) \mev$~\cite{Vlowk3N}.

\begin{figure}[t]
\includegraphics[width=0.46\textwidth,angle=0]{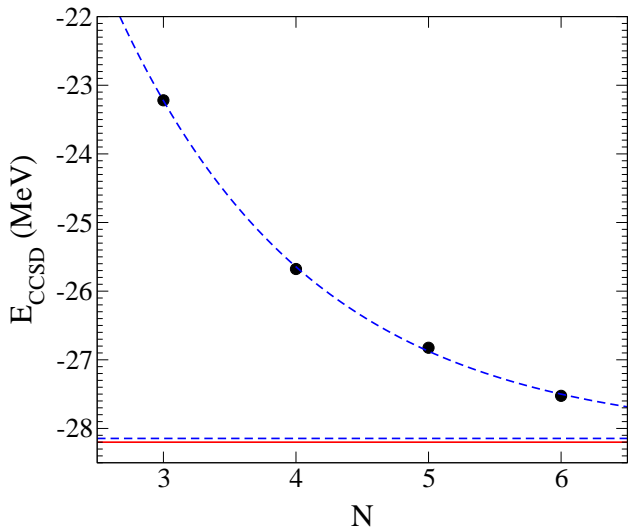}
\caption{(Color online) Data points: CCSD results 
(taken at the $\hbar \omega$ minima) for the binding energy 
of $^4$He with 3NFs as a function of the 
number of oscillator shells. Dashed lines: Exponential fit to 
the data and asymptote of the fit. Full line: Exact result.}
\label{fig5b}
\end{figure}

The coupled-cluster calculations are performed in a harmonic
oscillator (HO) basis, with basis parameters given by the oscillator
spacing $\hbar \omega$ and the number $N$ of oscillator shells. We
will present our results as a function of these parameters. The matrix
elements of the 3NF are calculated first in relative HO states given
by the expansion in 3N partial waves. For the transformation from
relative HO states to the single-particle (``$m$-scheme'') basis, we
essentially follow the Appendix of Ref.~\cite{Nog06}.  However, we can
also start from a 3NF in the non-antisymmetrized HO basis, which
simplifies the transformations given in Eqs.~(B.9), (B.11), and (B.12)
of Ref.~\cite{Nog06}, and explicitly antisymmetrize the $m$-scheme
matrix elements in the last step of our transformation. This approach
does not require coefficients of fractional parentage. We verified
that our matrix elements agree with those obtained
from a transformation based on antisymmetrized relative HO matrix
elements. We further have checked that the transformation preserves
the unit matrix and that it yields identical matrix elements for ${\bm
\sigma}_1 \cdot {\bm \sigma}_2$ and ${\bm \sigma}_2 \cdot {\bm
\sigma}_3$ in antisymmetrized states.

In Fig.~\ref{fig5}, our CCSD results for the binding energy of $^4$He
are shown with and without 3NFs as a function of the oscillator
spacing and with increasing model space size.  The 3NF contribution is
repulsive, in agreement with the Faddeev-Yakubovsky
calculation~\cite{Vlowk3N}, and correspondingly, the minima in $\hbar
\omega$ are shifted to smaller oscillator spacings.  We observe a slow
convergence at the last few $100 \kev$ level, which is due to the
sharp cutoff in $\vlowk$. This might be improved by using low-momentum
interactions with smooth cutoffs~\cite{Vlowksmooth}.  Using the minima
of the CCSD results with 3NFs, we make an exponential fit of the form
$E(N)=E_\infty + a \exp{(-b N)}$ to the data points. The result is
shown in Fig.~\ref{fig5b}.  The extrapolated infinite model space
value is $E_\infty = -28.09 \mev$, which is very close to the exact
result $E = -28.20(5) \mev$. 

\begin{figure}[t]
\includegraphics[width=0.46\textwidth,angle=0]{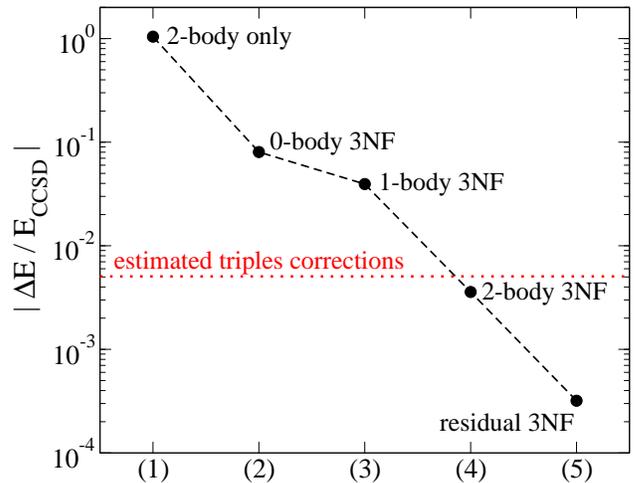}
\caption{(Color online) Relative contributions $|\Delta E / E|$ to 
the binding energy of $^4$He at the CCSD level. The different
points denote the contributions from 
(1) low-momentum NN interactions, 
(2) the vacuum expectation value of the 3NF, 
(3) the normal-ordered one-body Hamiltonian due to the 3NF,
(4) the normal-ordered two-body Hamiltonian due to the 3NF,
and (5) the residual 3NFs. The dotted line estimates the 
corrections due to omitted three-particle--three-hole clusters.}
\label{fig6}
\end{figure}

It is interesting to analyze the different contributions $\Delta E$ to
the binding energy $E$. The individual contributions are given in
Fig.~\ref{fig6} for a model space of $N=4$ oscillator shells and
$\hbar \omega = 20 \mev$. The main contribution stems from the
low-momentum NN interaction.  The contributions from 3NFs account only
for about 10\% of the total binding energy. This is consistent with
the chiral EFT power-counting estimate $\langle V_{\rm 3N} \rangle
\sim (Q/\Lambda_\chi)^3 \: \langle \vlowk \rangle \approx 0.1 \:
\langle \vlowk \rangle$~\cite{Vlowk3N} (see also Table~1 in
Ref.~\cite{Schw05}). The second, third, and fourth largest
contribution are due to the first, second, and third term on the
right-hand side of Eq.~(\ref{normal}). These are the density-dependent
zero-, one-, and two-body terms, which resulted from the normal
ordering of the three-body Hamiltonian in coupled-cluster theory.  The
contributions from the residual three-body Hamiltonian,
Eq.~(\ref{h3}), are very small and are represented by the last point
in Fig.~\ref{fig6}.  Recall that the residual 3NF contributes to the
energy directly through Eq.~(\ref{erg}) and indirectly through a
modification of the cluster amplitudes via Eqs.~(\ref{t1fac})
and~(\ref{t2fac}). Apparently, both contributions are very small.
In addition and independent of the result that low-momentum
3N interactions are perturbative for cutoffs $\Lambda \lesssim
2 \fmi$~\cite{Vlowk3N}, we find here that the contributions of
3NFs decrease rapidly with increasing rank of the normal-ordered terms.

\begin{figure}[t]
\includegraphics[width=0.46\textwidth,angle=0]{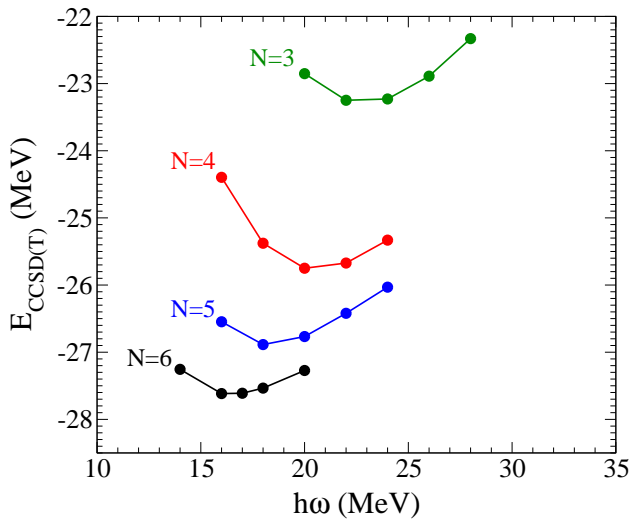}
\caption{(Color online) CCSD(T) results for the binding energy of
$^4$He as a function of the oscillator spacing and for model 
spaces consisting of $N=3$ to $N=6$ oscillator shells. The 
contributions from 3NFs are limited to the density-dependent zero-, 
one-, and two-body terms and exclude its residual three-body terms.}
\label{fig7}
\end{figure}

The small contribution from the residual three-body Hamiltonian is the
most important result of our study. It suggests that one can neglect
the residual terms of the 3NF when computing binding energies of light
nuclei. This is not unexpected and has been anticipated in several
earlier studies. Mihaila and Heisenberg~\cite{Mi00} computed the
charge form factor for $^{16}$O within coupled-cluster theory and
found a very good agreement with experimental data by considering only
the density-dependent one- and two-body parts of 3NFs. Similarly,
Navr{\'a}til and Ormand~\cite{NavOr} observed in no-core shell-model
calculations that density-dependent two-body terms are the most
significant contributions of effective three-body forces. Our finding
also support Zuker's~\cite{Zuker} idea that monopole corrections to
valence-shell interactions are due to the density-dependent terms of
3NFs. Note finally that the modeling of three-body interactions
in terms of density-dependent two-body Hamiltonians has a long
history, see e.g. Ref.~\cite{Negele}.  Note that all these examples and the
present study employ sufficiently "soft" or "effective" interactions.
We expect that the smallness of residual 3NFs is a property of such
interactions. We will study the cutoff dependence of this finding in
future work.  Finally, the smallness of residual 3NFs is also
encouraging for future improved nuclear matter calculations, which
currently include low-momentum 3NFs through density-dependent NN
interactions~\cite{nucmatt}.

The smallness of the residual three-body terms 
is also for coupled-cluster calculations a most welcome result.
This is attractive for two reasons. First, the inclusion
of the residual three-nucleon Hamiltonian, as described in
Subsection~\ref{maintheo}, is computationally expensive. It exceeds
the cost of a CCSD calculation for two-body Hamiltonians by a factor
of order $O(n_u) + O(n_o^2)$ and is therefore significant for a large
number of unoccupied orbitals and/or large number of nucleons.
Second, the omission of the residual three-body Hamiltonian will allow
us to treat 3NFs within the standard coupled-cluster theory developed
for two-body Hamiltonians (after normal ordering). As a result, we can
take the CCSD calculations one step further and include perturbative
corrections of three-particle--three-hole clusters~\cite{Deegan}.

\begin{figure}[t]
\includegraphics[width=0.46\textwidth,angle=0]{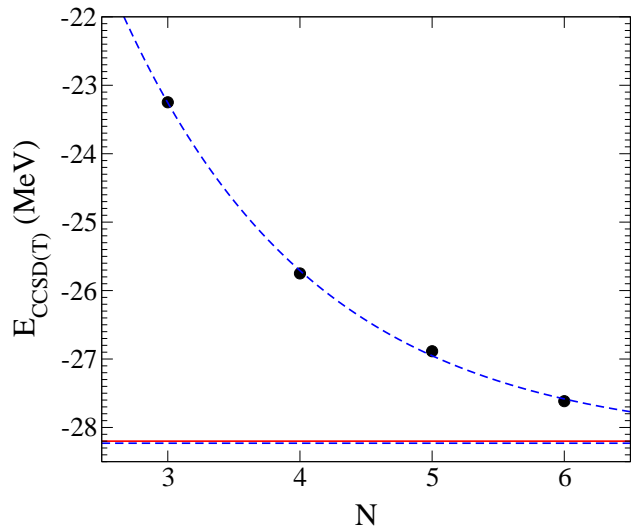}
\caption{(Color online) Data points: CCSD(T) results 
(taken at the $\hbar \omega$ minima) for the binding energy 
of $^4$He with 3NFs as a function of the 
number of oscillator shells. Dashed lines: Exponential fit to 
the data and asymptote of the fit. Full line: Exact result.}
\label{fig7b}
\end{figure}

Let us neglect the residual 3NF terms of Eq.~(\ref{h3})
and perform CCSD(T) calculations for the binding energy of $^4$He. The
approximate inclusion of three-particle--three-hole clusters improves
the accuracy of our calculations. Our results are shown in
Fig.~\ref{fig7}. The comparison with the CCSD full lines in
Fig.~\ref{fig5} shows that the triples corrections add about $100-200
\kev$ of additional binding energy (at the minimum) for fixed number
$N$ of oscillator shells, and somewhat weakens the $\hbar \omega$
dependence. 

An exponential extrapolation of the (approximate) CCSD(T) minima to an
infinite model space is shown in Fig.~\ref{fig7b} and yields $E_\infty
= -28.24 \mev$. This is in excellent agreement with the exact
Faddeev-Yakubovsky result $E=-28.20(5) \mev$. In our largest model
space at the minimum $\hbar\omega = 17 \mev$, the ground-state
expectation values for the center-of-mass Hamiltonian is $\langle
H_{\rm cm} \rangle \approx 20 \kev$ while the expectaion value for the
angular momentum is zero for a closed-shell nucleus by construction.
These results are very good, and it remains to be seen whether a more
sophisticated treatment of triples excitations~\cite{Pie05} would lead
to further improvements. We expect that the expectation value of the
center-of-mass Hamiltonian decreases with increasing size of the model
space.

Finally, we also show the size of the CCSD(T) corrections in
Fig.~\ref{fig6} as the horizontal dotted line.  Clearly, these
contributions are more important than the contributions from the
residual 3NF terms, and this observation fully justifies the 
omission of the latter.

\section{Summary}
\label{summary}

We have developed coupled-cluster theory for three-body Hamiltonians
in the two-particle--two-hole cluster approximation (CCSD). We derived 
the corresponding coupled-cluster equations directly in a factorized 
form and thereby avoided the explicit construction and analysis of a
considerable number of diagrams that enter these equations. The resulting 
formulae were used for a very efficient numerical implementation.

We have performed {\it ab-initio} coupled-cluster calculations
based on low-momentum NN and 3N interactions for the binding
energy of $^4$He and compared to the exact Faddeev-Yakubovsky
result. The 3NF contributions to the zero-, one-, and two-body 
terms of the normal-ordered Hamiltonian are dominant.
The contributions from residual 3NFs are smaller 
than the corrections due to three-particle--three-hole cluster 
excitations and can therefore be safely neglected. 
Future work will include all 3N partial waves and studies
of the cutoff dependence and of the convergence properties
using low-momentum interactions with smooth cutoffs. 
Our findings tremendously simplify the computational cost of 
coupled-cluster theory with 3NFs. This opens the avenue to explore 
3NFs in medium-mass nuclei and to investigate questions related to 
modern nuclear interactions. 

\section*{Acknowledgments}

We thank D.~Bernholdt, S.K.~Bogner, R.J.~Furnstahl and R.J.~Harrison
for useful discussions. This research was supported in part by the
Laboratory Directed Research and Development program of Oak Ridge
National Laboratory (ORNL), by the U.S. Department of Energy under
Contract Nos. \ DE-AC05-00OR22725 with UT-Battelle, LLC (ORNL), and
DE-FC02-07ER41457 (University of Washington), and under Grant Nos.\
DE-FG02-96ER40963 (University of Tennessee), DE-FG02-97ER41014
(University of Washington), DE-FG02-01ER15228 (Michigan State
University), and by the Natural Sciences and Engineering Research
Council of Canada (NSERC). TRIUMF receives federal funding via a
contribution agreement through the National Research Council of
Canada. Computational resources were provided by the National Center
for Computational Sciences at Oak Ridge, the National Energy Research
Scientific Computing Facility, and the John von Neumann Institute for
Computing in J\"ulich, Germany.

\end{document}